\documentclass[aps,pra, twocolumn, groupedaddress, twoside]{revtex4-2}

\usepackage{graphicx}
\usepackage{amsfonts}
\usepackage{amsmath}
\usepackage{amssymb}

\usepackage{xcolor}
\definecolor{grey}{rgb}{0.7,0.7,0.7}
\definecolor{db}{rgb}{0,0,0.5}

\usepackage{fancyhdr}
 
\begin{document}



\title{\colorbox{db}{\parbox{\linewidth}{  \centering \parbox{0.9\linewidth}{ \textbf{\centering\Large{\color{white}{ \vskip0.2em{ Generic Optical Singularities in Brewster-reflected Post-paraxial Beam-fields}\vskip0.8em}}}}}}}



\author{Anirban Debnath}
\email[]{anirban.debnath090@gmail.com}
\affiliation{School of Physics, University of Hyderabad, Hyderabad 500046, India}

\author{Nirmal K. Viswanathan}
\email[]{nirmalsp@uohyd.ac.in}
\affiliation{School of Physics, University of Hyderabad, Hyderabad 500046, India}


\date{\today}

\begin{abstract}
\noindent{\color{grey}{\rule{0.784\textwidth}{1pt}}}
\vspace{-0.8em}

Brewster-reflection of a post-paraxial optical beam at a plane dielectric interface unravels fundamentally significant optical singularity dynamics. We express the simulated field-component profiles of a Brewster-reflected post-paraxial beam-field via empirical functions, using which we demonstrate optical beam-shifts and formation of phase singularities. These occurrences naturally reveal the presence and complex transitional-dynamics of generic polarization-singularities --- which we observe via simulation and experiments. A single reflection being the core process, our method becomes a fundamentally appealing way to generate optical singularities and to study their dynamics. 
{\color{grey}{\rule{0.784\textwidth}{1pt}}}
\end{abstract}


\maketitle



\tableofcontents

{\color{grey}{\noindent\rule{\linewidth}{1pt}}}


\section{Introduction}

The strong influence of the optical fields carrying well-defined phase and polarization singularities has revolutionized the fundamental studies and applications of the wave-fields and their interactions with matter \cite{Gbur, NyeBerry1974, Bhandari97, SGVMH97, PA2000, SV2001, DOP09, BNRev, Nye83b, Nye83a, Hajnal87a, Hajnal87b, NH1987, DH1994, DennisPS02, DennisMonstar08, NKVMonstar, NKVFiber, Vpoint, WMCpoint}. 
In addition to their significant contribution towards the fundamental understanding of the electromagnetic characteristics of light, optical singularities provide core concepts in many emerging areas of research --- including Majorana and Poincarana sphere formalisms \cite{Hannay1998a, Hannay1998b, Poincarana}, topological and non-Hermitian photonics \cite{BerryNonH, LuTop, OzTop, FengNonH, GanNonH, MiriNonH, ChenNonH}, optical M\"obius strips \cite{FreundMob, BauerMob, BauerMob2, GalvezMob, EtxarriMob}, polarization knots \cite{BerryKnot, LeachKnot, DennisKnot, LarocqueKnot}, 
superoscillations \cite{BerrySO95, BerrySO09, ACSSO17, BerrySO19};
and are also directly related to several interdisciplinary research areas --- including crystal-dislocation, 
trapping and rotation of micro-particles, 
classical and quantum communications, microscopy and imaging 
etc. \cite{Gbur, Roadmap}.

Generic optical singularities are broadly classified into two fundamentally related but distinctly characterized categories \cite{Gbur}: (i) phase singularity; and (ii) polarization singularity.
Generic phase singularities are points (in 2D) or lines (in 3D) in space where both real and imaginary parts of a complex field become zero; thus making the phase of the field undefined \cite{NyeBerry1974, Bhandari97, SGVMH97, PA2000, SV2001, DOP09, BNRev, Gbur}. Polarization singularities are 
regions where specific properties of the polarization ellipse are undefined: orientation (for $C$-point); handedness (for $L$-line); both orientation and handedness (for $V$-point) \cite{Gbur, Nye83b, Nye83a, Hajnal87a, Hajnal87b, NH1987, DH1994, SV2001, DOP09, DennisPS02, DennisMonstar08, NKVMonstar, NKVFiber, Vpoint}. The polarization ellipses around these singularities organize themselves into unique patterns such as lemon, star, monstar, node, center, saddle etc. (topological indices respectively $+1/2$, $-1/2$, $-1/2$, $+1$, $+1$, $-1$) \cite{Gbur, Nye83b, Nye83a, Hajnal87a, Hajnal87b, NH1987, DH1994, SV2001, DOP09, DennisPS02, DennisMonstar08, NKVMonstar, NKVFiber, Vpoint}. Standard methods of generating vortex beams with embedded polarization singularities \cite{VortexBrewster, Cia2003, Marr06, ZaoSOI, Brass2009, BliokhSOI, Manni2011, APS12, NKVSagnac, NKVMonstar, EnZ, Gbur} thus involve, in some form or the other,
the inherent inhomogeneous polarization of the considered beam-field \cite{Bliokh2006, Bliokh2007, Dennis, Gotte, BARev, ADNKVarXiv, Serna09}.

A classic case of inhomogeneous polarization in the beam cross-section occurs for the reflection and transmission of a composite optical wave, where,
the constituent plane waves reflect and transmit differently due to differences in Fresnel coefficients \cite{SalehTeich, BornWolf, Jackson}; thus generating complicated reflected and transmitted field profiles \cite{GH, Artmann, Fedorov, Schilling, Imbert, Onoda, Bliokh2006, Bliokh2007, HostenKwiat, Dennis, Gotte, BARev, Berry435, ADNKVarXiv}. This effect is especially unique when a constituent wave with transverse magnetic (TM) polarization is incident at Brewster angle ($\theta_B$) \cite{SalehTeich, BornWolf, Jackson}, because this constituent wave is not reflected at all. Thus, a point of zero TM reflection is created in an otherwise-nonzero reflected field profile; indicating the existence of a phase singularity at that zero point. Also, the strongly inhomogeneous polarization around that zero point indicates the existence of polarization singularities.

The presence of phase singularity in a non-paraxial Brewster-reflected beam-field was recently observed by Barczyk et al. \cite{VortexBrewster}. However, they observed it in a specific experiment and in a complex setup --- 
thus, not explicitly identifying this phase singularity as a generic occurrence irrespective of the used setup.
On the other hand, in Ref. \cite{CLEO2020} we have reported some initial simulation-based observations of polarization singularities in a Brewster-reflected paraxial beam-field. But to our knowledge, a detailed analysis of the polarization singularities and their complex transitional-dynamics, appearing in a Brewster-reflected beam-field, is not currently present in literature.

We address the above-mentioned unexplored and fundamental electromagnetic-optical problems in the present work. We first simulate a Brewster-reflected post-paraxial optical beam-field. We then express this complete field in terms of empirical functions; and use them to demonstrate the occurrences of beam-shifts and phase singularities. 
Then we use the calculated information to perform a detailed analysis of the polarization singularities and their various complex transitional-dynamics in the beam-field. In particular, we demonstrate effects such as merger of two $C$-point singularities into a higher-order $V$-point singularity \cite{DH1994, Gbur}; pair-disappearance of singularities at infinity; and isolation of a $C$-point singularity for off-Brewster-angle reflection. 
Finally, we verify our analytical and simulated results via experimental observations.



\section{The Optical System}

\begin{figure}[t]
\includegraphics[width = 0.75\linewidth]{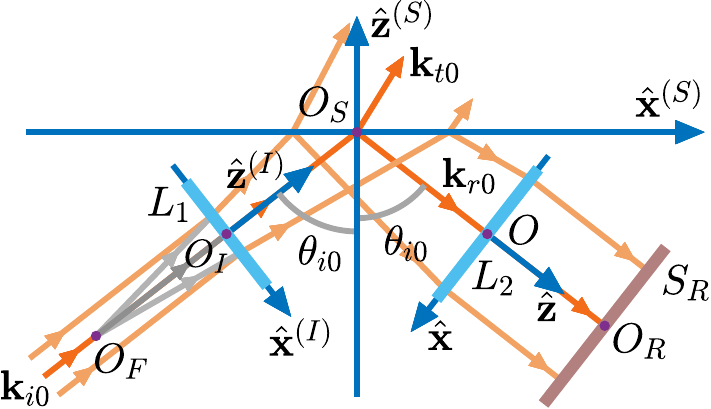} 
\caption{Schematic of the simulated optical system (description in the text).}{\color{grey}{\rule{\linewidth}{1pt}}}
\label{Fig_Setup}
\end{figure}

To explore Brewster-reflected beam-field profiles, we have simulated an optical system [FIG. \ref{Fig_Setup}] based on our formalism developed in Ref. \cite{ADNKVarXiv}.
In a medium of refractive index $n_1$, an initial collimated Gaussian beam is converted to a weakly-diverging post-paraxial Gaussian beam by passing it through a concave lens $L_1$ of focal length $\mathcal{F}_1 = - O_F O_I$. 
This beam is incident on an isotropic dielectric interface of refractive index $n_2$. The central angle of incidence is $\theta_{i0}$. The reflected beam cross-sectional profile is observed at the screen $S_R$ after collimating it through a convex lens $L_2$ of focal length $\mathcal{F}_2 = O_F O_I + O_I O_S + O_S O$. By setting $\theta_{i0} = \theta_{B}$, the Brewster-reflected beam-profile is observed.


\section{Analysis}\label{Sec_Analysis}

\begin{figure*}[t]
\includegraphics[width = \linewidth]{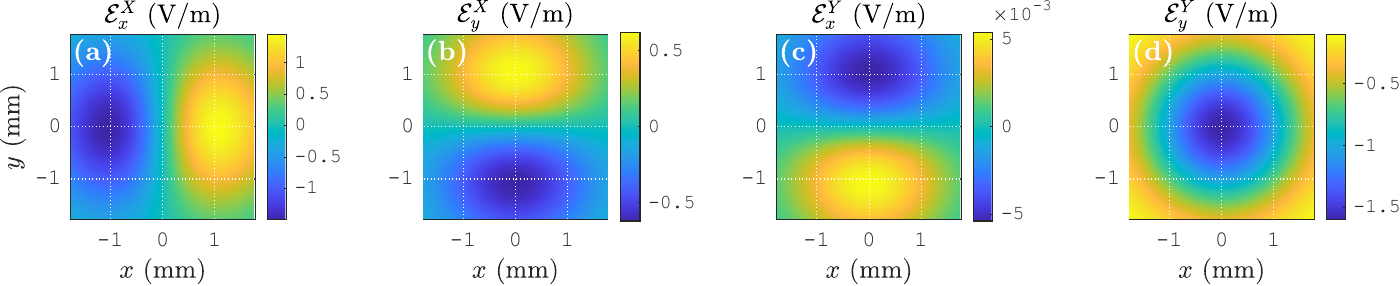} 
\caption{Simulated field-component profiles [Eqs. (\ref{E=EX+eEY}, \ref{EXY=ExXY+EyXY})] at the screen $S_R$ [FIG. \ref{Fig_Setup}] --- \textbf{(a)} $\mathcal{E}_x^X$, \textbf{(b)} $\mathcal{E}_y^X$, \textbf{(c)} $\mathcal{E}_x^Y$, \textbf{(d)} $\mathcal{E}_y^Y$ --- for the chosen system parameters and for $\theta_{i0} = \theta_B$, $\theta_E = 0.5^\circ$ [Eq. (\ref{E0xyI})]. The colorbars express the concerned functions in V/m.}
{\color{grey}{\rule{\linewidth}{1pt}}}
\label{Fig_ExyXYprofiles}
\end{figure*}

We define global incident and reflected beam coordinate systems $I(x^{(I)}, y^{(I)}, z^{(I)})$ and $R(x,y,z)$ with respect to the central incident and reflected wavevectors $\mathbf{k}_{i0}$ and $\mathbf{k}_{r0}$ [FIG. \ref{Fig_Setup}]. 
We define (i) the initial input collimated beam (before the lens $L_1$) in the $I$ coordinate system; and (ii) analyze the final output collimated beam (after the lens $L_2$) in the $R$ coordinate system. The incident diverging beam after $L_1$ and the reflected diverging beam before $L_2$ are analyzed in the simulation \citep{ADNKVarXiv} (including the lens transformations); whose mathematical details are not required here. The mathematical forms of only the (i) initial and (ii) final collimated beams are relevant for the purpose of the present paper. Both of these beams are approximated via plane wavefronts, which are characterized by conveniently omissible constant phase terms over the beam cross-sections. So, we analyze these plane-wave beam-fields only in terms of their complex field amplitude vectors, denoted by $\boldsymbol{\mathcal{E}}$.

In the $I$ coordinate system, the Gaussian electric field amplitude profile of the initial collimated beam is given by
\begin{eqnarray}
&\boldsymbol{\mathcal{E}}_0^{(I)} 
= \boldsymbol{\mathcal{E}}_{0x}^{(I)} + e^{i\Phi_E} \boldsymbol{\mathcal{E}}_{0y}^{(I)} = \mathcal{E}_{0x}^{(I)} \hat{\mathbf{x}}^{(I)} + e^{i\Phi_E} \mathcal{E}_{0y}^{(I)} \hat{\mathbf{y}}^{(I)}; \hspace{1em} & \label{E0I} 
\end{eqnarray}
\vspace{-2em}
\begin{equation}
\mathcal{E}_{0x}^{(I)} = \mathcal{E}_{00}\, e^{-\rho^{(I)\,2}/w_0^2} \cos\theta_E; \hspace{0.3em}
\mathcal{E}_{0y}^{(I)} = \mathcal{E}_{00}\, e^{-\rho^{(I)\,2}/w_0^2} \sin\theta_E;  \label{E0xyI}
\end{equation}
where, $\mathcal{E}_{00} $ is the electric field magnitude at the beam-axis; $\rho^{(I)} = \left( x^{(I)\,2} + y^{(I)\,2} \right)^{1/2}$; $w_0 $ is the half-beam-width; $\theta_E$, $\Phi_E $ are angle and relative-phase parameters which determine the exact polarization of $\boldsymbol{\mathcal{E}}_{0}^{(I)}$. After passing through the lens $L_1$, the beam-divergence is obtained as $2\theta_D = -2 \tan^{-1}(w_0/\mathcal{F}_1)$.

With respect to the plane of incidence, $\boldsymbol{\mathcal{E}}_{0x}^{(I)}$ is a TM polarized field and $\boldsymbol{\mathcal{E}}_{0y}^{(I)}$ is a transverse electric (TE) polarized field. The observed field $\boldsymbol{\mathcal{E}}$ at the screen $S_R$, though collimated by the lens $L_2$ to approximate a plane-wave beam-field, is neither TM nor TE --- because, in the intermediate stage, the reflection of the curved incident wavefront at the dielectric interface fundamentally induces polarization-inhomogeneity across the wavefront \cite{Bliokh2006, Bliokh2007, Dennis, Gotte, BARev, ADNKVarXiv, Serna09}. 
The final output field $\boldsymbol{\mathcal{E}}$ at the screen $S_R$ is obtained in the form
\begin{eqnarray}
& \boldsymbol{\mathcal{E}} = \boldsymbol{\mathcal{E}}^{X} + e^{i\Phi_E} \boldsymbol{\mathcal{E}}^{Y}; & \label{E=EX+eEY} \\ 
& 
\boldsymbol{\mathcal{E}}^{X} = \mathcal{E}_x^{X} \hat{\mathbf{x}} + \mathcal{E}_y^{X} \hat{\mathbf{y}}, \hspace{0.5em} 
\boldsymbol{\mathcal{E}}^{Y} = \mathcal{E}_x^{Y} \hat{\mathbf{x}} + \mathcal{E}_y^{Y} \hat{\mathbf{y}}; 
& \label{EXY=ExXY+EyXY} 
\end{eqnarray}
where, $\boldsymbol{\mathcal{E}}^{X}$ and $\boldsymbol{\mathcal{E}}^{Y}$ are the individual final output fields corresponding to $\boldsymbol{\mathcal{E}}_{0x}^{(I)}$ and $\boldsymbol{\mathcal{E}}_{0y}^{(I)}$ respectively.

Here onwards we set the following parameter values in the simulation:
free-space wavelength $\lambda = 632.8$ nm; refractive indices $n_1 = 1$, $n_2 = 1.52$ ($\theta_B = \tan^{-1}(n_2/n_1) \approx 56.66^\circ$); incident beam power $P_w = 1$ mW; 
$w_0 = 0.6$ mm; $\mathcal{F}_1 = -5$ cm (thus, $\theta_D \approx 0.7^\circ$); propagation distances $O_I O_S = 5$ cm, $O_S O = 2.5$ cm; $\mathcal{F}_2 = 12.5$ cm. 
Using the above-mentioned values of $P_w$ and $w_0$ in Eq. (\ref{E00=E00(Pw)}), we obtain the central field-magnitude $\mathcal{E}_{00} \approx 1.15430 \times 10^3$ V/m.

For incidence around Brewster angle, the Fresnel TM reflection coefficient is very less than the Fresnel TE reflection coefficient \cite{SalehTeich, BornWolf, Jackson}. So, if $|\boldsymbol{\mathcal{E}}_{0x}^{(I)}| \sim |\boldsymbol{\mathcal{E}}_{0y}^{(I)}|$, then we get $|\boldsymbol{\mathcal{E}}^{Y}| \gg |\boldsymbol{\mathcal{E}}^{X}|$; and the subtle effects such as the ones reported here become obscured due to the dominance of $\boldsymbol{\mathcal{E}}^{Y}$. It is thus essential to have $|\boldsymbol{\mathcal{E}}^{Y}| \sim |\boldsymbol{\mathcal{E}}^{X}|$, which can be achieved by appropriately choosing $\theta_E$ in Eq. (\ref{E0xyI}). Here onwards, we choose $\theta_E = 0.5^\circ$ for most of our general observations for $\theta_{i0} = \theta_B$. We also choose $\theta_E = 0^\circ$ in some cases to observe only $\boldsymbol{\mathcal{E}}^{X}$ for Brewster reflection.

For $\theta_{i0} = \theta_B$ and $\theta_E = 0.5^\circ$, the simulated field-component profiles $\mathcal{E}_x^{X}$, $\mathcal{E}_y^{X}$, $\mathcal{E}_x^{Y}$ and $\mathcal{E}_y^{Y}$ at the surface $S_R$, as functions of coordinates $(x,y)$, are shown in FIG. \ref{Fig_ExyXYprofiles}.
These profiles 
show that the functions $\mathcal{E}_x^{X}$, $\mathcal{E}_y^{X}$, $\mathcal{E}_x^{Y}$ and $\mathcal{E}_y^{Y}$ can be expressed in empirical forms (2D Hermite-Gaussian functions \cite{GriffithsQM, SalehTeich}) as
\begin{eqnarray}
&\mathcal{E}_x^{X} = A_1 f \cos\phi; \hspace{1em} \mathcal{E}_y^{X} = A_2 f \sin\phi; & \label{ExXEyX=A1A2}\\
&\mathcal{E}_x^{Y} = -A_3 f \sin\phi; \hspace{1em} \mathcal{E}_y^{Y} = -A_4\, g; & \label{ExYEyY=A3A4}\\
& \mbox{where,} \hspace{1em} g = e^{-\rho^2/w_R^2}; \hspace{1em} f = (\rho/w_R) g; & \label{fg_def}
\end{eqnarray}
where, $(\rho,\phi)$ are the plane polar coordinates corresponding to $(x,y)$; and $A_1$, $A_2$, $A_3$, $A_4$, $w_R$ are constants which are determined numerically. For the specific example of FIG. \ref{Fig_ExyXYprofiles}, these numerically determined values are $A_1 \approx 3.42226$ V/m, $A_2 \approx 1.44286$ V/m, $A_3 \approx 1.25917 \times 10^{-2}$ V/m, $A_4 \approx 1.59497$ V/m, $w_R \approx 1.49603$ mm; which give less than $0.02$ fractional error in the maximum magnitudes of the considered functions as compared to the exact field-component profiles obtained via simulation. For $\theta_E = 0^\circ$, we automatically get $A_3 = A_4 = 0$ V/m; 
numerically obtain $A_1 \approx 3.42239$ V/m, $A_2 \approx 1.44292$ V/m; with $w_R \approx 1.49603$ mm remaining unchanged.

Physically, the constant $w_R$ is a half-beam-width term corresponding to the profiles of FIG. \ref{Fig_ExyXYprofiles}, as understood from the functional form of $g$ [Eq. (\ref{fg_def})]. The constants $A_j$ ($j = 1,2,3,4$) are field magnitude terms. Each $A_j$ can be expressed in the form $A_j = a_j\, \mathcal{E}_{00}$; where $a_j$ is a dimensionless quantity giving the ratio $A_j/\mathcal{E}_{00}$. For example, as per the calculated value of $\mathcal{E}_{00}$ and the numerically obtained values of $A_j$ for the case of FIG. \ref{Fig_ExyXYprofiles}, the ratios are $a_1 \approx 2.96479 \times 10^{-3}$, $a_2 \approx 1.24999 \times 10^{-3}$, $a_3 \approx 1.09085 \times 10^{-5}$, $a_4 \approx 1.38176 \times 10^{-3}$. These ratios signify the fractional contributions of $\mathcal{E}_{00}$ in the formation of the field profiles $\mathcal{E}_x^X$, $\mathcal{E}_y^X$, $\mathcal{E}_x^Y$, $\mathcal{E}_y^Y$ via Eqs. (\ref{ExXEyX=A1A2}, \ref{ExYEyY=A3A4}). The ratios are as small as $\sim 10^{-3}$ and $\sim 10^{-5}$; signifying the very small reflected beam-intensity as compared to the incident beam-intensity under the concerned Brewster-incidence and near-Brewster-incidence conditions.

For the purpose of the present paper, it is sufficient to proceed further by considering the empirical functions of Eqs. (\ref{ExXEyX=A1A2}--\ref{ExYEyY=A3A4}). Our complete mathematical analysis on the exact field $\boldsymbol{\mathcal{E}}$ [Eq. (\ref{E=EX+eEY})], based on the reflection and transmission coefficient matrix formalism \citep{ADNKVarXiv}, will be reported elsewhere.


We use Eqs. (\ref{EXY=ExXY+EyXY}--\ref{ExYEyY=A3A4}) in Eq. (\ref{E=EX+eEY}) to find the empirical form of the field $\boldsymbol{\mathcal{E}}$;
and then express it in terms of the $\hat{\boldsymbol{\sigma}}^\pm$ spin-states as
\begin{eqnarray}
& \hspace{-1em} \boldsymbol{\mathcal{E}} = \mathcal{E}_+ \hat{\boldsymbol{\sigma}}^+ + \mathcal{E}_- \hat{\boldsymbol{\sigma}}^-; \hspace{1em} \mbox{where,} \hspace{1em} \mathcal{E}_\pm = a_\pm + i b_\pm; & \label{E=SUMsigmapm,Epm=a+ib} \\
& \hspace{-1em} \mbox{\small{ $ a_\pm = \dfrac{g}{\sqrt{2}} \left[ \left( A_1 \dfrac{x}{w_R} - A_3 \dfrac{y}{w_R} \cos\Phi_E \right) \mp A_4 \sin\Phi_E \right]; $ } } & \label{a_pm} \\
& \hspace{-1em} \mbox{\small{ $ b_\pm = \dfrac{g}{\sqrt{2}} \left[ -A_3 \dfrac{y}{w_R} \sin\Phi_E \mp \left( A_2 \dfrac{y}{w_R} - A_4 \cos\Phi_E \right) \right]; $ } } & \label{b_pm}
\end{eqnarray}
where, $x = \rho \cos\phi$, $y = \rho \sin\phi$ are used. 
The intensities $I$, $I_\pm$ of the fields $\boldsymbol{\mathcal{E}}$, $\mathcal{E}_\pm \hat{\boldsymbol{\sigma}}^\pm$ are then obtained in the form $I = (n_1 / 2\mu_0 c) |\boldsymbol{\mathcal{E}}|^2$ as
\begin{eqnarray}
\hspace{-1.5em} I &=& \frac{n_1}{2\mu_0 c} \, g^2 \left[ A_1^2 \frac{x^2}{w_R^2} + \left(A_2^2 + A_3^2\right) \frac{y^2}{w_R^2} \right. \nonumber\\ && \left. - 2 \left( A_1 A_3 \frac{xy}{w_R^2} + A_2 A_4 \frac{y}{w_R} \right) \cos\Phi_E + A_4^2 \right]. \label{I=main} \\
\hspace{-1.5em} I_\pm &=& \frac{I}{2} \mp \frac{n_1}{2\mu_0 c} \, g^2 \left( A_1 A_4 \frac{x}{w_R} - A_2 A_3 \frac{y^2}{w_R^2} \right) \sin\Phi_E. \label{I_pm}
\end{eqnarray}
Thus, the fields $\mathcal{E}_\pm \hat{\boldsymbol{\sigma}}^\pm$ have different intensities $I_\pm$, but satisfy energy conservation $I_+ + I_- = I$. 
Equations (\ref{E=SUMsigmapm,Epm=a+ib}--\ref{I_pm}) are the primary corollaries of our central assumptions [Eq. (\ref{ExXEyX=A1A2}--\ref{fg_def})]. Based on these, the secondary corollaries such as the total powers, centroid positions and singularity positions are derived.

\vspace{0.5em}

\textbf{Powers:} 
From $I$, $I_\pm$, we obtain the corresponding powers $P$, $P_\pm$ in the form $P = \int_{-\infty}^\infty \int_{-\infty}^\infty I \, dx \, dy$ as
\begin{eqnarray}
& P = \dfrac{n_1}{2\mu_0 c} \dfrac{\pi w_R^2}{8} A_0^2; & \label{P=main} \\
& P_\pm = \dfrac{P}{2} \pm \dfrac{n_1}{2\mu_0 c} \dfrac{\pi w_R^2}{8} A_2 A_3 \sin\Phi_E; & \label{P_pm} \\
&\mbox{where,} \hspace{1em} A_0 = \left( A_1^2 + A_2^2 + A_3^2 + 4 A_4^2 \right)^{\frac{1}{2}}. & \label{A0}
\end{eqnarray}
Here, $P_+ \neq P_-$ because $I_+ \neq I_-$ [Eq. (\ref{I_pm})].
However, as per the numerically obtained constant values, $A_3 \ll A_0$. So the power difference is very small.

\vspace{0.5em}

\textbf{Centroids and Beam Shifts:} 
The centroid positions $(x_C, y_C)$, $(x_{C\pm}, y_{C\pm})$ of the intensity profiles $I$, $I_\pm$ are obtained in the form
$$ x_C = \dfrac{1}{P} \int_{-\infty}^\infty \int_{-\infty}^\infty x I \, dx \, dy; \hspace{0.5em} y_C = \dfrac{1}{P} \int_{-\infty}^\infty \int_{-\infty}^\infty y I \, dx \, dy.
$$
We thus get
\begin{eqnarray}
& x_C = 0; \hspace{1em} y_C = -2 w_R \dfrac{A_2 A_4}{A_0^2} \cos\Phi_E; & \label{xC,yC}\\
& x_{C\pm} = \mp \dfrac{2 w_R A_1 A_4 \sin\Phi_E}{A_0^2 \pm 2 A_2 A_3 \sin\Phi_E}; & \label{xCpm_main}\\
& y_{C\pm} = -\dfrac{2 w_R A_2 A_4 \cos\Phi_E}{A_0^2 \pm 2 A_2 A_3 \sin\Phi_E}. & \label{yCpm_main}
\end{eqnarray}

Zero $x_C$ [Eq. (\ref{xC,yC})] implies zero GH shift \cite{GH, Artmann, Bliokh2006, Bliokh2007, Dennis, Gotte, BARev} of the field $\boldsymbol{\mathcal{E}}$ \cite{FOOTNOTE_GH}.
However, non-zero IF shift \cite{Fedorov, Schilling, Imbert, Bliokh2006, Bliokh2007, Dennis, Gotte, BARev} $y_C$ [Eq. (\ref{xC,yC})] of $\boldsymbol{\mathcal{E}}$; longitudinal and transverse spin-shifts $x_{C\pm}$ [Eq. (\ref{xCpm_main})] and $y_{C\pm}$ [Eq. (\ref{yCpm_main})] (i.e. $x$-shifts and $y$-shifts of $\mathcal{E}_\pm\hat{\boldsymbol{\sigma}}^\pm$) are observed. 

Since $A_2, A_4 \sim A_0$, Eq. (\ref{xC,yC}) shows that $|y_C| \sim w_R$ for appropriate ranges of $\Phi_E$. Thus, for such $\Phi_E$ ranges, giant IF shift of the order of the half-beam-width $w_R$ is observed. This result has been experimentally demonstrated by G\"otte et al. \cite{GotteLofflerDennis}.

A graphical description of the origin of IF shift is obtained from the profiles of FIG. \ref{Fig_ExyXYprofiles}. The $\mathcal{E}_y^Y$ profile is negative everywhere; whereas, the $\mathcal{E}_y^X$ profile is positive in the $y > 0$ region and negative in the $y < 0$ region. So, for $-\pi/2 < \Phi_E < \pi/2$, these two fields undergo destructive interference in the $y > 0$ region and constructive interference in the $y < 0$ region. The opposite happens for $-\pi < \Phi_E < -\pi/2$ and $\pi/2 < \Phi_E \leq \pi$.
This gives rise to an asymmetry in the total intensity profile $I$ about the $x$ axis; thus resulting in the IF shift. Though we have demonstrated this effect in the specific case of Brewster reflection, we have verified via simulation that these behavior and explanation are valid for any angle of incidence.

Now, with the approximation $A_3 \ll A_0$, we rewrite $(x_{C\pm},y_{C\pm})$ [Eqs. (\ref{xCpm_main}, \ref{yCpm_main})] as
\begin{eqnarray}
& x_{C\pm} = 2 w_R \dfrac{A_1 A_4}{A_0^2} \sin\Phi_E \left( 2 \dfrac{A_2 A_3}{A_0^2} \sin\Phi_E \mp 1 \right); & \label{xCpm_approx} \\
& y_{C\pm} = y_C + \delta_\pm; & \label{yCpm_approx} \\
& \mbox{where,} \hspace{0.5em} \delta_\pm = \pm 4 w_R \dfrac{A_2^2 A_3 A_4}{A_0^4} \cos\Phi_E \sin\Phi_E. & \label{deltapm}
\end{eqnarray}
The first term in the parentheses of Eq. (\ref{xCpm_approx}) is small, since $A_3 \ll A_0$. So, for appropriate ranges of $\Phi_E$, the $x_{C\pm}$ shifts are large (i.e. comparable to $w_R$) and are approximately opposite to each other. Thus, while Qin et al. \cite{Qin2011} report wavelength-order in-plane spin-shifts for non-Brewster-reflection cases, the present Brewster-reflection case shows `giant' (beam-size-order) longitudinal spin-shifts. 

The transverse spin-shifts $y_{C\pm}$ [Eq. (\ref{yCpm_approx})] are the IF shift $y_C$ [Eq. (\ref{xC,yC})] with added correction terms $\delta_\pm$ [Eq. (\ref{deltapm})]. These $\delta_\pm$ terms are the spin Hall shifts \cite{Bliokh2006, Liberman, Onoda, HostenKwiat, XieSHELinIF}. The effect that the spin Hall shifts in Brewster reflection appear on both sides of IF shift has been demonstrated by Xie et al. \cite{XieSHELinIF}.

\begin{figure*}[t]
\includegraphics[width = \linewidth]{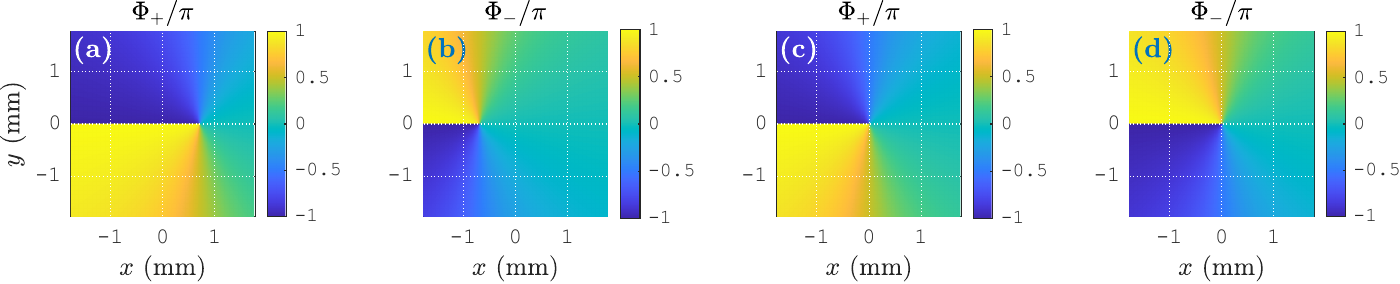} 
\caption{\textbf{(a, b)} Simulated $\Phi_\pm$ phase profiles 
[Eq. (\ref{Phi_pm})] of the fields $\mathcal{E}_\pm\hat{\boldsymbol{\sigma}}^\pm$ [Eq. (\ref{E=SUMsigmapm,Epm=a+ib})] for $\theta_{i0} = \theta_B$, $\theta_E = 0.5^\circ$, $\Phi_E = \pi/2$; showing $l^{\mp}$ phase singularities at $(x_{S\pm},y_{S\pm})$ [Eqs. (\ref{xSpm_main}--\ref{ySpm_approx})]. \textbf{(c, d)} Similar 
$\Phi_\pm$ 
profiles for 
$\theta_{i0} = \theta_B$, $\theta_E = 0^\circ$; with phase-singularity coordinates $x_{S\pm} = y_{S\pm} = 0$.}
{\color{grey}{\rule{\linewidth}{1pt}}}
\label{Fig_PhaseVortex}
\end{figure*}

\vspace{0.5em}

\textbf{Phase Singularities:} 
From Eq. (\ref{E=SUMsigmapm,Epm=a+ib}) we obtain the phases of $\mathcal{E}_\pm \hat{\boldsymbol{\sigma}}^\pm$ as
\begin{equation}
\Phi_\pm = \tan^{-1}(b_\pm / a_\pm). \label{Phi_pm}
\end{equation}
The phase-singularities of $\mathcal{E}_\pm \hat{\boldsymbol{\sigma}}^\pm$ appear where $a_\pm = b_\pm = 0$.
Applying this condition to the expressions of $a_\pm$, $b_\pm$ [Eqs. (\ref{a_pm}, \ref{b_pm})], we obtain the singularity coordinates $(x_{S\pm},y_{S\pm})$ as
\begin{eqnarray}
& x_{S\pm} = \dfrac{w_R A_4 (A_3 \pm A_2 \sin\Phi_E)}{A_1 (A_2 \pm A_3 \sin\Phi_E)}; & \label{xSpm_main} \\
& y_{S\pm} = \dfrac{w_R A_4 \cos\Phi_E}{A_2 \pm A_3 \sin\Phi_E}; & \label{ySpm_main} 
\end{eqnarray}
which we rewrite by using the approximation $A_3 \ll A_1, A_2, A_4$ as 
\begin{eqnarray}
& x_{S\pm} = w_R \dfrac{A_3 A_4}{A_1 A_2} \cos^2\Phi_E \pm w_R \dfrac{A_4}{A_1} \sin\Phi_E; & \label{xSpm_approx} \\
& y_{S\pm} = w_R \dfrac{A_4}{A_2} \cos\Phi_E \mp w_R \dfrac{A_3 A_4}{A_2^2} \cos\Phi_E \sin\Phi_E. & \label{ySpm_approx}
\end{eqnarray}
The first term in $x_{S\pm}$ [Eq. (\ref{xSpm_approx})] is small, since $A_3$ is small. So, the second term dominates for appropriate ranges of $\Phi_E$; and hence, for those $\Phi_E$ ranges, $x_{S+}$ and $x_{S-}$ are almost opposite to each other and also $|x_{S\pm}| \sim w_R$. However, for other appropriate $\Phi_E$ ranges, the first term in $y_{S\pm}$ [Eq. (\ref{ySpm_approx})] dominates while the second term is small due to the presence of $A_3$. Thus, the difference between $y_{S+}$ and $y_{S-}$ is small as compared to $y_{S\pm}$. 

The simulated $\Phi_+$ profile [Eq. (\ref{Phi_pm})] for $\theta_{i0} = \theta_B$, $\theta_E = 0.5^\circ$, $\Phi_E = \pi/2$ [FIG. \ref{Fig_PhaseVortex}(a)] shows 
an $l^-$ (i.e. topological charge $l = -1$ \cite{Gbur}) phase singularity at $(x_{S+},y_{S+})$. The $\Phi_-$ profile for the same simulation parameters [FIG. \ref{Fig_PhaseVortex}(b)]
shows an $l^+$ (i.e. topological charge $l = +1$) phase singularity at $(x_{S-},y_{S-})$. 
These singularities and the associated helical phase profiles indicate that the $\mathcal{E}_\pm \hat{\boldsymbol{\sigma}}^\pm$ fields carry orbital angular momenta (OAM) \cite{Gbur}. 

\vspace{0.5em}

\textbf{Special Case:} 
For the special case of $\theta_E = 0^\circ$, we get $\boldsymbol{\mathcal{E}}_0^{(I)} = \boldsymbol{\mathcal{E}}_{0x}^{(I)}$ [Eq. (\ref{E0I}, \ref{E0xyI})], and hence $\boldsymbol{\mathcal{E}} = \boldsymbol{\mathcal{E}}^X$ [Eq. (\ref{E=EX+eEY})]. All $A_3$ and $A_4$ terms in Eqs. (\ref{ExYEyY=A3A4}--\ref{ySpm_approx}) are zero in this case, and hence several simplified symmetries in the fields are observed: $I_+ = I_- = I/2$ 
[Eq. (\ref{I=main}, \ref{I_pm})]; $P_+ = P_- = P/2$ [Eq. (\ref{P=main}, \ref{P_pm})]; 
$y_C = 0$ [Eq. (\ref{xC,yC})], implying zero IF shift of the $\boldsymbol{\mathcal{E}}$ field; $x_{C\pm} = y_{C\pm} = 0$ [Eqs. (\ref{xCpm_approx}, \ref{yCpm_approx})], implying zero spin-shifts of both $\mathcal{E}_\pm\hat{\boldsymbol{\sigma}}^\pm$ fields; $x_{S\pm} = y_{S\pm} = 0$ [Eqs. (\ref{xSpm_approx}, \ref{ySpm_approx})], implying singularity positions at the screen-center $O_R$ [FIG. \ref{Fig_Setup}] for both $\mathcal{E}_\pm\hat{\boldsymbol{\sigma}}^\pm$ fields [FIGs. \ref{Fig_PhaseVortex}(c, d)].




\section{Polarization Singularities}

\begin{figure*}[t]
\includegraphics[width = \linewidth]{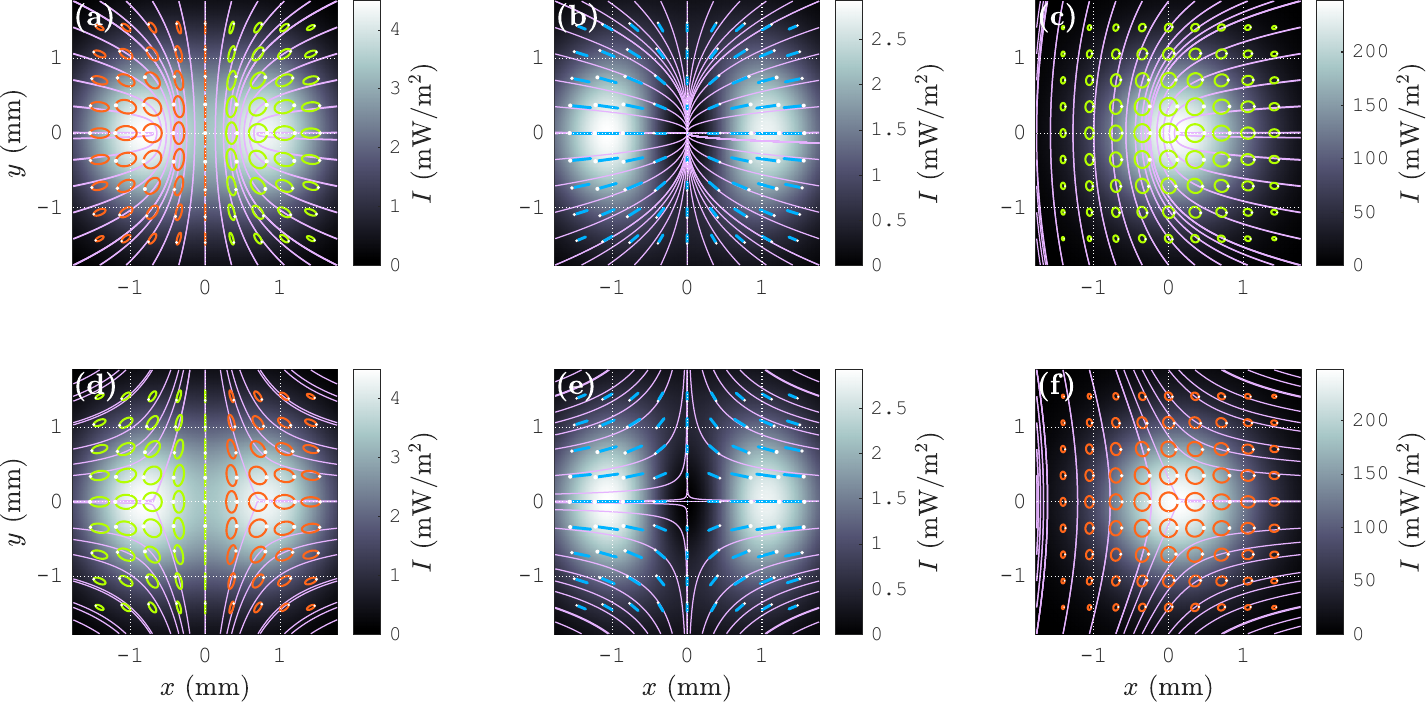}
\caption{Simulated profiles: 
\textbf{(a)} Complete field $\boldsymbol{\mathcal{E}}$ [Eq. (\ref{E=EX+eEY})] for $\theta_{i0} = \theta_B$, $\theta_E = 0.5^\circ$, $\Phi_E = \pi/2$; showing two $C$-point singularities (double lemon pattern) separated by an $L$-line singularity. 
\textbf{(b)} Field $\boldsymbol{\mathcal{E}}^X$ [Eq. (\ref{EXY=ExXY+EyXY})], obtained from the field $\boldsymbol{\mathcal{E}}$ of FIG. \ref{Fig_Phi+_DLemon_LemonStar}(a) by reducing $\theta_E$ to $0^\circ$. The field $\boldsymbol{\mathcal{E}}^X$ contains a node singularity. 
\textbf{(c)} Complete field $\boldsymbol{\mathcal{E}}$ for $\Delta \theta_{i0} = -2.0^\circ$, $\theta_E = 3.2^\circ$, $\Phi_E = \pi/2$; showing an isolated $C$-singularity (single lemon pattern). 
\textbf{(d)} The double star pattern, \textbf{(e)} the saddle singularity and \textbf{(f)} the single star pattern, created by passing the fields of FIGs. \ref{Fig_Phi+_DLemon_LemonStar}(a, b, c) respectively through a HWP (fast axis along $\hat{\mathbf{x}}$ or $\hat{\mathbf{y}}$). 
The field-representation color codes are: Light Green: LEP; Dark Orange: REP; Light Blue: Linear polarization; White Dot: Represents the tip of the field-vector $\boldsymbol{\mathcal{E}}$, originated at the ellipse-center, at time $t = 0$ (the complete description of this color-code representation is given in Ref. \cite{ADNKVarXiv}). The intensity profile $I$ [Eq. (\ref{I=main})] of the field $\boldsymbol{\mathcal{E}}$ is shown at the background of each $\boldsymbol{\mathcal{E}}$ profile. 
}
{\color{grey}{\rule{\linewidth}{1pt}}}
\label{Fig_Phi+_DLemon_LemonStar}
\end{figure*}

Equations (\ref{xCpm_approx}, \ref{xSpm_approx}) show that $x_{C\pm}$ and $x_{S\pm}$ are in directions opposite to each other. For example, 
for $\Phi_E = \pi/2$, we get 
$x_{C+}, x_{S-} \sim -w_R$ and $x_{C-}, x_{S+} \sim w_R$.
Thus, in the $x < 0$ region, the $\mathcal{E}_+ \hat{\boldsymbol{\sigma}}^+$ field with an almost constant phase profile is superposed on the relatively weak $\mathcal{E}_- \hat{\boldsymbol{\sigma}}^-$ field with an $l^+$ phase singularity; 
whereas, in the $x>0$ region, the $\mathcal{E}_- \hat{\boldsymbol{\sigma}}^-$ field with an almost constant phase profile is superposed on the relatively weak $\mathcal{E}_+ \hat{\boldsymbol{\sigma}}^+$ field with an $l^-$ phase singularity. 
These $\left(l^\pm \hat{\boldsymbol{\sigma}}^\mp + l^0 \hat{\boldsymbol{\sigma}}^\pm\right)$ superpositions around the phase singularities generate lemon polarization patterns 
in the complete field $\boldsymbol{\mathcal{E}}$ \cite{NKVMonstar}. The resulting $\boldsymbol{\mathcal{E}}$ field profile for $\Phi_E = \pi/2$ is shown in FIG. \ref{Fig_Phi+_DLemon_LemonStar}(a); where, two opposite lemon patterns around $C$-point singularities, symmetrically separated by an $L$-line singularity, are observed.

The other fundamental polarization pattern around a $C$-point singularity is the star pattern, 
which is obtained via $\left(l^\pm \hat{\boldsymbol{\sigma}}^\pm + l^0 \hat{\boldsymbol{\sigma}}^\mp\right)$ superpositions 
\cite{NKVMonstar}. 
As understood from Section \ref{Sec_Analysis}, 
the $l^\pm \hat{\boldsymbol{\sigma}}^\pm$ combinations do not appear automatically in the $\boldsymbol{\mathcal{E}}$ field, and hence the automatic appearance of the star pattern in Brewster reflection is prohibited. However, by passing the $\boldsymbol{\mathcal{E}}$ beam-field through a half wave plate (HWP), its $\left(l^\pm \hat{\boldsymbol{\sigma}}^\mp + l^0 \hat{\boldsymbol{\sigma}}^\pm\right)$ superpositions can be converted to $\left(l^\pm \hat{\boldsymbol{\sigma}}^\pm + l^0 \hat{\boldsymbol{\sigma}}^\mp\right)$ superpositions --- thus creating star patterns. The double lemon pattern of FIG. \ref{Fig_Phi+_DLemon_LemonStar}(a) is thus converted to the double star pattern of FIG. \ref{Fig_Phi+_DLemon_LemonStar}(d) with either $\hat{\mathbf{x}}$ or $\hat{\mathbf{y}}$ orientation of the HWP's fast axis.

The simulated $\boldsymbol{\mathcal{E}}^X$ field profile, obtained for $\theta_E = 0^\circ$, is shown in FIG. \ref{Fig_Phi+_DLemon_LemonStar}(b). It contains a node singularity, 
whose formation can be described in terms of a singularity-merger phenomenon \cite{DH1994, Gbur}. As the angle $\theta_E$ [Eq. (\ref{E0xyI})] is gradually changed from a non-zero value to $0^\circ$, the contribution of $\boldsymbol{\mathcal{E}}^Y$ in the total field $\boldsymbol{\mathcal{E}}$ [Eq. (\ref{E=EX+eEY})] is gradually reduced. Consequently, $|x_{S\pm}|$ and $|y_{S\pm}|$ [Eqs. (\ref{xSpm_approx}, \ref{ySpm_approx})] gradually reduce to zero. The phase profiles of FIGs. \ref{Fig_PhaseVortex}(a, b) thus gradually transform to the phase profiles of FIGs. \ref{Fig_PhaseVortex}(c, d) respectively; and correspondingly, the two lemon $C$-points of FIG. \ref{Fig_Phi+_DLemon_LemonStar}(a) approach each other to eventually merge at the screen-center $O_R$. This explains the formation of the node singularity of FIG. \ref{Fig_Phi+_DLemon_LemonStar}(b) for $\theta_E = 0^\circ$. 
The streamlines in FIG. \ref{Fig_Phi+_DLemon_LemonStar}(b) are not radial because of the complicated nature of the exact contributions of $\mathcal{E}_x^X \hat{\mathbf{x}}$ and $\mathcal{E}_y^X \hat{\mathbf{y}}$ [Eqs. (\ref{EXY=ExXY+EyXY}, \ref{ExXEyX=A1A2})] in the field $\boldsymbol{\mathcal{E}}^X$, as understood from FIGs. \ref{Fig_ExyXYprofiles}(a, b).

Observed through a HWP (fast axis along $\hat{\mathbf{x}}$ or $\hat{\mathbf{y}}$), the above-mentioned merger phenomenon is equivalent to the merger of the two star $C$-points of FIG. \ref{Fig_Phi+_DLemon_LemonStar}(d) to form the saddle singularity 
\cite{DH1994, Gbur} of FIG. \ref{Fig_Phi+_DLemon_LemonStar}(e).

Though we emphasize on considering small $\theta_E$ values giving $|\boldsymbol{\mathcal{E}}^{Y}| \sim |\boldsymbol{\mathcal{E}}^{X}|$, it is also essential to observe that, as $\theta_E$ is gradually increased to $90^\circ$, the total field $\boldsymbol{\mathcal{E}}$ gradually transforms to only $\boldsymbol{\mathcal{E}}^Y$ [Eqs. (\ref{E0I}--\ref{E=EX+eEY})] --- a non-singular field [FIGs. \ref{Fig_ExyXYprofiles}(c, d)]. 
Correspondingly, in terms of polarization singularities, $|x_{S\pm}|$ and $|y_{S\pm}|$ [Eqs. (\ref{xSpm_approx}, \ref{ySpm_approx})] gradually increase (depending on $\Phi_E$). The two $C$-lemons of FIG. \ref{Fig_Phi+_DLemon_LemonStar}(a) thus gradually move farther apart; and eventually disappear at infinity for $\theta_E = 90^\circ$. This process, though not equivalent, is comparable to the disappearance of phase vortices at infinity due to free-space propagation \cite{SGVMH97, Gbur}.


Manipulation of our setup based on the above 
understanding enables us to generate various other polarization-singular patterns. For instance, the symmetry of Brewster reflection can be broken by taking $\theta_{i0}$ away from $\theta_B$. For $\theta_{i0} = \theta_B + \Delta\theta_{i0}$, where $|\Delta\theta_{i0}| \gtrsim \theta_D$ and $|\Delta\theta_{i0}| \ll \theta_B$, the $\boldsymbol{\mathcal{E}}$ field-profile takes the form of either the right or the left lemon pattern of FIG. \ref{Fig_Phi+_DLemon_LemonStar}(a). Such a lemon pattern for $\Delta\theta_{i0} = -2.0^\circ$, $\theta_E = 3.2^\circ$, $\Phi_E = \pi/2$ is shown in FIG. \ref{Fig_Phi+_DLemon_LemonStar}(c). 
This single lemon pattern is converted to the single star pattern of FIG. \ref{Fig_Phi+_DLemon_LemonStar}(f) by passing the beam-field through a HWP (fast axis along $\hat{\mathbf{x}}$ or $\hat{\mathbf{y}}$).

\section{Experimental Observations}

\begin{figure}
\includegraphics[width = \linewidth]{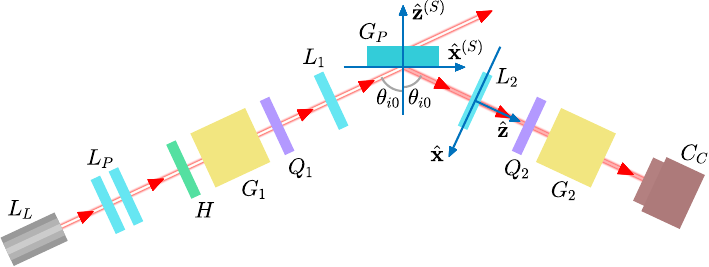}
\caption{The experimental setup. Here,  $L_L : $ He-Ne laser; $L_P : $ Collimating lens pair; $H : $ Half wave plate; $G_1, G_2 : $ Glan-Thompson polarizers; $Q_1, Q_2 : $ Quarter wave plates; $L_1 : $ Concave lens of focal length $\mathcal{F}_1 = -5$ cm [FIG. \ref{Fig_Setup}]; $L_2 : $ Convex lens of focal length $\mathcal{F}_2 = 12.5$ cm [FIG. \ref{Fig_Setup}];
$G_P : $ Glass plate, whose $-\hat{\mathbf{z}}^{(S)}$-facing surface is used as the concerned dielectric interface [FIG. \ref{Fig_Setup}]; $C_C : $ CCD camera, used as the screen $S_R$ [FIG. \ref{Fig_Setup}].}
{\color{grey}{\rule{\linewidth}{1pt}}}
\label{Fig_ExpSetup}
\end{figure}

\begin{figure*}[t]
\includegraphics[width = \linewidth]{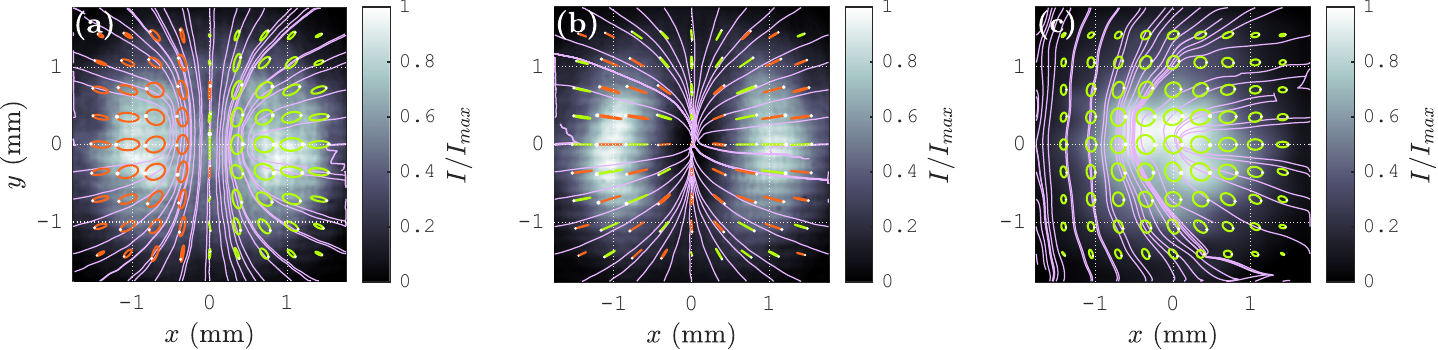}
\caption{\textbf{(a, b, c)} Experimental field profiles corresponding to the simulated profiles of FIGs. \ref{Fig_Phi+_DLemon_LemonStar}(a, b, c) respectively. 
The intensity profiles in the background are expressed in normalized form, $I/I_{max}$.}
{\color{grey}{\rule{\linewidth}{1pt}}}
\label{Fig_Experimental}
\end{figure*}

\begin{figure}[t]
\includegraphics[width = \linewidth]{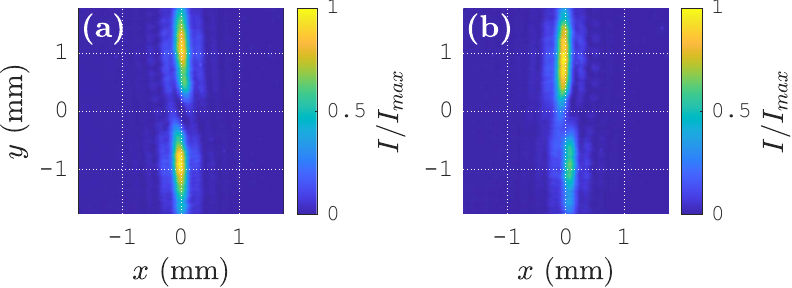} 
\caption{Experimental single-slit diffraction patterns of the fields \textbf{(a)} $\mathcal{E}_+\hat{\mathbf{d}}^+$ and \textbf{(b)} $\mathcal{E}_-\hat{\mathbf{d}}^-$ [Eq. (\ref{EQ=Epdp+Emdm})]; showing $l^\mp$ topological charges. 
}
{\color{grey}{\rule{\linewidth}{1pt}}}
\label{Fig_SingleSlit}
\end{figure}

The fundamental phase and polarization (lemons, node) singularity patterns, simulated based on our formalism of Ref. \citep{ADNKVarXiv} applied to a Brewster-reflected post-paraxial optical beam, are also experimentally observed \cite{FOOTNOTE_Exp}.
The schematic of the experimental setup is shown in FIG. \ref{Fig_ExpSetup}. A He-Ne laser beam ($\lambda = 632.8$ nm) is first collimated using a lens pair. 
The desired polarization-ellipticity of this collimated beam is then realized by using a half-wave-plate--Glan-Thompson-polarizer--quarter-wave-plate (HWP-GTP-QWP) combination.
The resultant elliptically polarized plane-wave beam is then used as the input $\mathbf{k}_{i0}$--$\boldsymbol{\mathcal{E}}_0^{(I)}$ beam in a schematic identical to FIG. \ref{Fig_Setup}, with identical parameter values used in the simulations (as given in Section \ref{Sec_Analysis}). 
A surface of a glass plate 
is used as the concerned dielectric interface.
A CCD camera is used as the screen $S_R$ to observe the collimated beam-field $\boldsymbol{\mathcal{E}}$ after the lens $L_2$. The state of polarization of this final beam-field is measured via Stokes parameters method \cite{Goldstein} by using a QWP-GTP combination positioned in front of the camera. 

The experimentally generated $\boldsymbol{\mathcal{E}}$ profile [Eqs. (\ref{E=EX+eEY}, \ref{E=SUMsigmapm,Epm=a+ib})] for $\theta_{i0} = \theta_B$, $\theta_E = 0.5^\circ$, $\Phi_E = \pi/2$ is shown in FIG. \ref{Fig_Experimental}(a);
which transforms to the profile of FIG. \ref{Fig_Experimental}(b) for $\theta_E = 0^\circ$, i.e. for the $\boldsymbol{\mathcal{E}} = \boldsymbol{\mathcal{E}}^X$ case [Eq. (\ref{E=EX+eEY})]. The total field profile for the off-Brewster-incidence case of $\Delta\theta_{i0} = -2^\circ$, $\theta_E = 3.2^\circ$, $\Phi_E = \pi/2$ is shown in FIG. \ref{Fig_Experimental}(c).
The experimentally obtained field profiles of FIGs. \ref{Fig_Experimental}(a, b, c) match well with the corresponding simulation-generated profiles of FIGs. \ref{Fig_Phi+_DLemon_LemonStar}(a, b, c) respectively --- thus verifying the formation of generic optical singularities in the Brewster-reflected post-paraxial beam-field.

To observe the underlying topological charges in the $\mathcal{E}_\pm \hat{\boldsymbol{\sigma}}^\pm$ fields [Eq. (\ref{E=SUMsigmapm,Epm=a+ib})] for the $\theta_E = 0^\circ$ case, we first pass the output beam through an appropriately oriented QWP to transform the $\boldsymbol{\mathcal{E}}$ beam-field to 
\begin{equation}
\boldsymbol{\mathcal{E}}_Q = \mathcal{E}_+ \, \hat{\mathbf{d}}^+ + \mathcal{E}_- \, \hat{\mathbf{d}}^-; \label{EQ=Epdp+Emdm}
\end{equation}
where, $\hat{\mathbf{d}}^\pm = (\hat{\mathbf{x}} \pm \hat{\mathbf{y}})/\sqrt{2}$. 
The constituent $\mathcal{E}_\pm \, \hat{\mathbf{d}}^\pm$ 
beam-fields are then independently extracted by using a GTP with appropriate orientations. The individual $\mathcal{E}_\pm \, \hat{\mathbf{d}}^\pm$ beams are then passed through a single vertical slit (width $ = 0.55$ mm), placed along $x = 0$. 
The far-field diffraction patterns given by $\mathcal{E}_+ \, \hat{\mathbf{d}}^+$ and $\mathcal{E}_- \, \hat{\mathbf{d}}^-$ are shown in FIGs. \ref{Fig_SingleSlit}(a, b) respectively.
These patterns are the characteristic single-slit diffraction patterns of beams with $l^\mp$ topological charges \cite{SingleSlit, BekshaevRev2011}. Hence, these patterns directly show the presence of helical phases with $l^\mp$ topological charges in the $\mathcal{E}_\pm \, \hat{\mathbf{d}}^\pm$ fields, and hence in the $\mathcal{E}_\pm  \hat{\boldsymbol{\sigma}}^\pm$ component beam-fields. 

\section{Conclusion}

Even if standard methods \cite{VortexBrewster, Cia2003, Marr06, ZaoSOI, Brass2009, BliokhSOI, Manni2011, APS12, NKVSagnac, NKVMonstar, Vpoint, EnZ, Gbur} utilize special optical elements and setups to generate well-refined optical singularities, the singularity-generation itself is a remarkably fundamental phenomenon. 
We have demonstrated the formation of generic optical singularities due to Brewster-reflection of a post-paraxial beam-field. The core process in this singularity-generation is a single reflection at a plane dielectric interface --- one of the most elementary field-transformations in optics. The complex transitional dynamics of these singularities are demonstrated via simple modifications of the input beam parameters. 
The formation and transitional-dynamics of these generic optical singularities are thus clear demonstrations of the significant polarization-inhomogeneity \cite{Bliokh2006, Bliokh2007, Dennis, Gotte, BARev, ADNKVarXiv, Serna09} of the Brewster-reflected post-paraxial beam-field.
Hence, the present work serves the purpose of initiating a potential series of research works exploring the novel polarization patterns in a reflected beam-field under Brewster-incidence and near-Brewster-incidence conditions. Our subsequent works in this direction will be reported elsewhere.



From a practical perspective, the simplicity of our singularity-generation process comes at the cost of losing most of the beam-intensity in the transmitted beam [FIGs. \ref{Fig_Setup}, \ref{Fig_ExpSetup}], as understood from the numerically obtained ratios $a_j = A_j/\mathcal{E}_{00}$ [Section \ref{Sec_Analysis}]. Also, the generated beam is not in a pure LG mode \cite{SalehTeich, DOP09, Gbur}. However, our clear aim here is to explore the optical singularities in a Brewster-reflected post-paraxial beam-field; rather than to use the process as a singularity-generation technique. The above-mentioned inconveniences do not impact at all in analyzing the generic properties of the generated singularities; because, as long as the singularities exist, their generic properties also exist irrespective of the beam intensity and exact mode-composition \cite{Gbur}. 
Nevertheless, due to the simplicity of the process, one can also use it to readily generate optical singularities within any optical setup as necessary.

\section*{A\lowercase{ppendix}: D\lowercase{etermination of the} C\lowercase{entral} F\lowercase{ield-Magnitude of the} I\lowercase{nitial} I\lowercase{nput} B\lowercase{eam}}

In this section, we determine the central field-magnitude $\mathcal{E}_{00}$ of the initial input beam-field $\boldsymbol{\mathcal{E}}_0^{(I)}$ [Eq. (\ref{E0I})] from the total beam-power $P_w$. By using Eqs. (\ref{E0I}, \ref{E0xyI}), we express the field $\boldsymbol{\mathcal{E}}_0^{(I)}$ as
\begin{equation}
\boldsymbol{\mathcal{E}}_0^{(I)} = \mathcal{E}_{00} \, e^{-\rho^{(I)\, 2}/w_0^2} \left( \cos\theta_E \, \hat{\mathbf{x}}^{(I)} + e^{i\Phi_E} \sin\theta_E \, \hat{\mathbf{y}}^{(I)} \right).
\end{equation}
The corresponding magnetic field is
\begin{eqnarray}
\hspace{-4em}\boldsymbol{\mathcal{H}}_0^{(I)} &=& \left( \mathbf{k}_{i0} \times \boldsymbol{\mathcal{E}}_0^{(I)} \right) / \omega \mu_0 \nonumber \\
&=& \frac{n_1}{\mu_0 c} \, \mathcal{E}_{00} \, e^{-\rho^{(I)\, 2}/w_0^2} \left( -e^{i\Phi_E} \sin\theta_E \, \hat{\mathbf{x}}^{(I)} \right. \nonumber\\
&& \hspace{11em} \left. 
+ \cos\theta_E \, \hat{\mathbf{y}}^{(I)} \right);
\end{eqnarray}
where, $\mu_0 = $ magnetic permeability of free space; $c = $ speed of light in free space. The time-averaged Poynting vector is then given by
\begin{equation}
\langle \mathbf{S} \rangle = \frac{1}{2}\, \mathfrak{Re} \left( \boldsymbol{\mathcal{E}}_0^{(I) \, *} \times \boldsymbol{\mathcal{H}}_0^{(I)} \right) = \frac{n_1}{2 \mu_0 c} \, \mathcal{E}_{00}^2 \, e^{-2\rho^{(I)\, 2}/w_0^2} \, \hat{\mathbf{z}}^{(I)}.
\end{equation}
Then, the time-averaged beam power passing through the $z^{(I)} = 0$ plane is obtained as
\begin{equation}
P_w = \int_{-\infty}^{\infty} \int_{-\infty}^{\infty} \langle \mathbf{S} \rangle \cdot \hat{\mathbf{z}}^{(I)} dx^{(I)} dy^{(I)} 
= \frac{n_1}{2 \mu_0 c} \, \mathcal{E}_{00}^2 \, \frac{\pi w_0^2}{2}. \label{Pw=Pw(E00)}
\end{equation}
Thus, the power $P_w$ of the initial input Gaussian beam is equivalent to the power of a hypothetical plane-wave-beam having a
wavevector $\mathbf{k}_{i0}$ and a uniform field-magnitude $\mathcal{E}_{00}$ 
over a circular cross-section of radius $w_0/\sqrt{2}$.

In an actual experiment, the power $P_w$ and the half-beam-width $w_0$ of the input Gaussian beam are known. Using these known quantities, the central field-magnitude $\mathcal{E}_{00}$ is obtained by rearranging Eq. (\ref{Pw=Pw(E00)}) as
\begin{equation}
\mathcal{E}_{00} = \frac{2}{w_0} \left( \frac{\mu_0 c}{n_1 \pi} P_w \right)^\frac{1}{2}. \label{E00=E00(Pw)}
\end{equation}



\begin{acknowledgments}
We thank Ms. Shivangi Dubey and Mr. Mohamed Yaseen, NIT Warangal, India, for their 
help with the experimental works. A.D. thanks CSIR for research fellowship (JRF). N.K.V. thanks SERB for financial support.
\end{acknowledgments}



\begin{thebibliography}{99}



\bibitem{Gbur}
G. J. Gbur, \textit{Singular Optics} (CRC Press, Taylor \& Francis Group, LLC, FL, 2017).

\bibitem{NyeBerry1974}
J. F. Nye and M. V. Berry, Proc. R. Soc. Lond. A \textbf{336}, 165 (1974).

\bibitem{Bhandari97}
R. Bhandari, Phys. Rep. \textbf{281}, 1 (1997).

\bibitem{SGVMH97}
M. S. Soskin, V. N. Gorshkov, M. V. Vasnetsov, J. T. Malos, and N. R. Heckenberg, Phys. Rev. A \textbf{56}, 4064 (1997).

\bibitem{PA2000}
M. Padgett and A. Allen, Contemp. Phys. \textbf{41}, 275 (2000).

\bibitem{SV2001}
M. S. Soskin and M. V. Vasnetsov, Prog. Opt. \textbf{42}, 219 (2001).

\bibitem{DOP09}
M. R. Dennis, K. O'Holleran, and M. J. Padgett, Prog. Opt. \textbf{53}, 293 (2009).

\bibitem{BNRev}
K. Y. Bliokh and F. Nori, Phys. Rep. \textbf{592}, 1 (2015).

\bibitem{Nye83b}
J. F. Nye, Proc. R. Soc. Lond. A \textbf{387}, 105 (1983).

\bibitem{Nye83a}
J. F. Nye, Proc. R. Soc. Lond. A \textbf{389}, 279 (1983).

\bibitem{Hajnal87a}
J. V. Hajnal, Proc. R. Soc. Lond. A \textbf{414}, 433 (1987).

\bibitem{Hajnal87b}
J. V. Hajnal, Proc. R. Soc. Lond. A \textbf{414}, 447 (1987).

\bibitem{NH1987}
J. F. Nye and J. V. Hajnal, Proc. R. Soc. Lond. A \textbf{409}, 21 (1987).

\bibitem{DH1994}
T. Delmarcelle and L. Hesselink, in \textit{VIS '94: Proceedings of the Conference on Visualization '94} (IEEE, 1994) pp. 140--147.

\bibitem{DennisPS02}
M. R. Dennis, Opt. Commun. \textbf{213}, 201 (2002).

\bibitem{DennisMonstar08}
M. R. Dennis, Opt. Lett. \textbf{33}, 2572 (2008).

\bibitem{NKVMonstar}
N. K. Viswanathan, V. Kumar, and G. M. Philip, J. Opt. \textbf{15}, 044027 (2013).

\bibitem{NKVFiber}
Y. V. Jayasurya, V. V. G. K. Inavalli, and N. K. Viswanathan, Appl. Opt. \textbf{50}, E131 (2011).

\bibitem{Vpoint}
Ruchi, S. K. Pal, and P. Senthilkumaran, Opt. Express \textbf{25}, 19326 (2017).

\bibitem{WMCpoint}
S. Nechayev, M. Neugebauer, M. Vorndran, G. Leuchs, and P. Banzer, Phys. Rev. Lett. \textbf{121}, 243903 (2018).

\bibitem{Hannay1998a}
J. H. Hannay, J. Mod. Opt. \textbf{45}, 1001 (1998).

\bibitem{Hannay1998b}
J. H. Hannay, J. Phys. A: Math. Gen. \textbf{31}, L53 (1998).

\bibitem{Poincarana}
K. Y. Bliokh, M. A. Alonso, and M. R. Dennis, Rep. Prog. Phys. \textbf{82}, 122401 (2019).

\bibitem{BerryNonH}
M. V. Berry, Czechoslov. J. Phys. \textbf{54}, 1039 (2004).

\bibitem{LuTop}
L. Lu, J. D. Joannopoulos, and M. Solja\v{c}i\'c, Nat. Photonics \textbf{8}, 821 (2014).

\bibitem{OzTop}
T. Ozawa et al., Rev. Mod. Phys. \textbf{91}, 015006 (2019).

\bibitem{FengNonH}
L. Feng, R. El-Ganainy, and L. Ge, Nat. Photonics \textbf{11}, 752 (2017).

\bibitem{GanNonH}
R. El-Ganainy, K. G. Makris, M. Khajavikhan, Z. H. Musslimani, S. Rotter, and D. N. Christodoulides, Nat. Phys. \textbf{14}, 11 (2018).

\bibitem{MiriNonH}
M.-A. Miri and A. Al\`u, Science \textbf{363}, eaar7709 (2019).

\bibitem{ChenNonH}
W. Chen, Q. Yang, Y. Chen, and W. Liu, arXiv:2006.06517v2 [physics.optics] (2020).

\bibitem{FreundMob}
I. Freund, Opt. Commun. \textbf{283}, 1 (2010).

\bibitem{BauerMob}
T. Bauer, P. Banzer, E. Karimi, S. Orlov, A. Rubano, L. Marrucci, E. Santamato, R. W. Boyd, and G. Leuchs, Science \textbf{347}, 964 (2015).

\bibitem{BauerMob2}
T. Bauer, M. Neugebauer, G. Leuchs, and P. Banzer, Phys. Rev. Lett. \textbf{117}, 013601 (2016).

\bibitem{GalvezMob}
E. J. Galvez, I. Dutta, K. Beach, J. J. Zeosky, J. A. Jones, and B. Khajavi, Sci. Rep. \textbf{7}, 13653 (2017).

\bibitem{EtxarriMob}
A. Garcia-Etxarri, ACS Photonics \textbf{4}, 1159 (2017).

\bibitem{BerryKnot}
M. V. Berry and M. R. Dennis, Proc. R. Soc. A \textbf{457}, 2251 (2001).

\bibitem{LeachKnot}
J. Leach, M. R. Dennis, J. Courtial, and M. J. Padgett, Nature \textbf{432}, 165 (2004).

\bibitem{DennisKnot}
M. R. Dennis, R. P. King, B. Jack, K. O'Holleran, and M. J. Padgett, Nat. Phys. \textbf{6}, 118 (2010).

\bibitem{LarocqueKnot}
H. Larocque, D. Sugic, D. Mortimer, A. J. Taylor, R. Fickler, R. W. Boyd, M. R. Dennis, and E. Karimi, Nat. Phys. \textbf{14}, 1079 (2018).

\bibitem{BerrySO95}
M. V. Berry, in \textit{Quantum Coherence and Reality: In Celebration of the 60th Birthday of Yakir Aharonov}, edited by J. S. Anandan and J. L. Safko (World Scientific, Singapore, 1994) pp. 55--65.

\bibitem{BerrySO09}
M. V. Berry and M. R. Dennis, J. Phys. A: Math. Theor. \textbf{42}, 022003 (2009).

\bibitem{ACSSO17}
Y. Aharonov, F. Colombo, I. Sabadini, D. Struppa, and J. Tollaksen, Mem. Am. Math. Soc. \textbf{247}, 1 (2017).

\bibitem{BerrySO19}
M. V. Berry and P. Shukla, J. Opt. \textbf{21}, 064002 (2019).

\bibitem{Roadmap}
Rubinsztein-Dunlop et al., J. Opt. \textbf{19}, 013001 (2017).

\bibitem{VortexBrewster}
R. Barczyk, S. Nechayev, M. A. Butt, G. Leuchs, and P. Banzer, Phys. Rev. A \textbf{99}, 063820 (2019).

\bibitem{Cia2003}
A. Ciattoni, G. Cincotti, and C. Palma, J. Opt. Soc. Am. A \textbf{20}, 163 (2003).

\bibitem{Marr06}
L. Marrucci, C. Manzo, and D. Paparo, Phys. Rev. Lett. \textbf{96}, 163905 (2006).


\bibitem{ZaoSOI}
Y. Zhao, J. S. Edgar, G. D. M. Jeffries, D. McGloin, and D. T. Chiu, Phys. Rev. Lett. \textbf{99}, 073901 (2007).

\bibitem{Brass2009}
E. Brasselet, Y. Izdebskaya, V. Shvedov, A. S. Desyatnikov, W. Krolikowski, and Y. S. Kivshar, Opt. Lett. \textbf{34}, 1021 (2009).

\bibitem{BliokhSOI}
K. Y. Bliokh, E. A. Ostrovskaya, M. A. Alonso, O. G. Rodr\'iguez-Herrera, D. Lara, and C. Dainty, Opt. Express \textbf{19}, 26132 (2011).

\bibitem{Manni2011}
F. Manni et al., Phys. Rev. B \textbf{83}, 241307(R) (2011).

\bibitem{APS12}
M. Alonso, G. Piquero, and J. Serna, Opt. Commun. \textbf{285}, 1631 (2012).

\bibitem{NKVSagnac}
D. N. Naik and N. K. Viswanathan, J. Opt. \textbf{18}, 095601 (2016).

\bibitem{EnZ}
A. Ciattoni, A. Marini, and C. Rizza, Phys. Rev. Lett. \textbf{118}, 104301 (2017).

\bibitem{Bliokh2006}
K. Y. Bliokh and Y. P. Bliokh, Phys. Rev. Lett. \textbf{96}, 073903 (2006).

\bibitem{Bliokh2007}
K. Y. Bliokh and Y. P. Bliokh, Phys. Rev. E \textbf{75}, 066609 (2007).

\bibitem{Dennis}
M. R. Dennis and J. B. G\"otte, New J. Phys. \textbf{14}, 073013 (2012).

\bibitem{Gotte}
J. B. G\"otte and M. R. Dennis, New J. Phys. \textbf{14}, 073016 (2012).

\bibitem{BARev}
K. Y. Bliokh and A. Aiello, J. Opt. \textbf{15}, 014001 (2013).

\bibitem{ADNKVarXiv}
A. Debnath and N. K. Viswanathan, J. Opt. Soc. Am. A \textbf{37}, 1971 (2020).

\bibitem{Serna09}
J. Serna and G. Piquero, Opt. Commun. \textbf{282}, 1973 (2009).

\bibitem{SalehTeich}
B. E. A. Saleh and M. C. Teich, \textit{Fundamentals of Photonics}, 2nd ed. (John Wiley \& Sons, Inc., NJ, 2007).

\bibitem{BornWolf}
M. Born and E. Wolf, \textit{Principles of Optics}, 7th ed. (Cambridge University Press, Cambridge, 1999).

\bibitem{Jackson}
J. D. Jackson, \textit{Classical Electrodynamics}, 3rd ed. (John Wiley \& Sons (Asia) Pte. Ltd, Singapore, 1999).

\bibitem{GH}
V. F. Goos and H. H\"anchen, Ann. Physik \textbf{436}, 333 (1947).

\bibitem{Artmann}
K. Artmann, Ann. Physik \textbf{437}, 87 (1948).


\bibitem{Fedorov}
F. I. Fedorov, Dokl. Akad. Nauk SSSR \textbf{105}, 465
(1955).

\bibitem{Schilling}
H. Schilling, Ann. Physik \textbf{471}, 122 (1965).

\bibitem{Imbert}
C. Imbert, Phys. Rev. D \textbf{5}, 787 (1972).

\bibitem{Onoda}
M. Onoda, S. Murakami, and N. Nagaosa, Phys. Rev. Lett. \textbf{93}, 083901 (2004).

\bibitem{HostenKwiat}
O. Hosten and P. Kwiat, Science \textbf{319}, 787 (2008).

\bibitem{Berry435}
M. V. Berry, Proc. R. Soc. A \textbf{467}, 2500 (2011).

\bibitem{CLEO2020}
A. Debnath and N. K. Viswanathan, in \textit{Conference on Lasers and Electro-Optics} (Optical Society of America, 2020) p. JTh2E.1.

\bibitem{GriffithsQM}
D. J. Griffiths, \textit{Introduction to Quantum Mechanics}, 2nd ed. (Pearson Education, Inc., NJ, 2005).

\bibitem{GotteLofflerDennis}
J. B. G\"otte, W. L\"offler, and M. R. Dennis, Phys. Rev. Lett. \textbf{112}, 233901 (2014).

\bibitem{Qin2011}
Y. Qin, Y. Li, X. Feng, Y.-F. Xiao, H. Yang, and Q. Gong, Opt. Express \textbf{19}, 9636 (2011).

\bibitem{Liberman}
V. S. Liberman and B. Y. Zel'dovich, Phys. Rev. A \textbf{46},
5199 (1992).

\bibitem{XieSHELinIF}
L. Xie et al., Opt. Express \textbf{26}, 22934 (2018).

\bibitem{Goldstein}
D. H. Goldstein, \textit{Polarized Light}, 3rd ed. (CRC Press, Taylor \& Francis Group, FL, 2011).

\bibitem{SingleSlit}
D. P. Ghai, P. Senthilkumaran, and R. S. Sirohi, Opt Laser Eng \textbf{47}, 123 (2009).

\bibitem{BekshaevRev2011}
A. Bekshaev, K. Y. Bliokh, and M. Soskin, J. Opt. \textbf{13}, 053001 (2011).

\bibitem{FOOTNOTE_GH}
\textbf{Note:} This result is, however, based on the empirical functions of Eqs. (5, 6). We have observed a small correction (small with respect to $w_R$) to the GH shift by using the exact simulated $\boldsymbol{\mathcal{E}}$ field.

\bibitem{FOOTNOTE_Exp}
\textbf{Note:} It is sufficient to experimentally observe the profiles of FIGs. 4(a, b, c) (as in FIG. 6) to verify the formation of optical singularities in the Brewster-reflected post-paraxial beam-field. Though the profiles of FIGs. 4(d, e, f) explore additional features (HWP-transformations) of the beam-field, their explicit experimental verification is trivial and is not required for the purpose of the present paper.




\end{thebibliography}

\end{document}